\documentclass[twocolumn]{aastex63}

\usepackage[utf8]{inputenc} 
\usepackage[T1]{fontenc}    
\usepackage{hyperref}       
\usepackage{url}            
\usepackage{booktabs}       
\usepackage{amsmath,amsfonts,amssymb}
\usepackage{nicefrac}       
\usepackage{microtype}      
\usepackage{xspace}
\usepackage{natbib}
\usepackage{graphicx}

\newcommand{\zsq}{\ensuremath{Z^2_n}\xspace}
\newcommand{\zsqbin}{\ensuremath{Z^2_{n,\mathrm{bin}}}\xspace}
\newcommand{\zsqgauss}{\ensuremath{Z_{n, \mathrm{gauss}}}\xspace}
\newcommand{\ef}{\ensuremath{\mathcal{S}}\xspace}
\newcommand{\efbin}{\ensuremath{\mathcal{S}_{bin}}\xspace}
\newcommand{\nbin}{\ensuremath{N_{\rm bin}}\xspace}
\newcommand{\nfmpb}{\ensuremath{N_\mathrm{fmpb}}\xspace}
\newcommand{\nfmpbi}{\ensuremath{N_{\mathrm{fmpb}, i}}\xspace}
\newcommand{\ssig}{\ensuremath{S_{\rm sig}}\xspace}
\newcommand{\sthr}{\ensuremath{S_{\rm thr}}\xspace}
\newcommand{\dof}{\ensuremath{\nu}\xspace}
\newcommand{\chizsq}{\ensuremath{\chi^2_{2n}}\xspace}

\renewcommand{\eqref}{Eq.~\ref}
\newcommand{\figref}{Figure~\ref}
\newcommand{\secref}{Section~\ref}

\received{June 1, 2019}
\revised{January 10, 2019}
\accepted{\today}
\submitjournal{ApJ}

\shorttitle{Generalized $Z^2_n$ and $H$ statistics}
\shortauthors{Bachetti et al.}

\graphicspath{{./}{figures/}}

\begin{document}

\title{Extending the $Z^2_n$ and $H$ statistics to generic pulsed profiles}

\author[0000-0002-4576-9337]{Matteo Bachetti}
\affiliation{INAF-Osservatorio Astronomico di Cagliari, via della Scienza 5, I-09047 Selargius (CA), Italy}

\author[0000-0001-7397-8091]{Maura Pilia}
\affiliation{INAF-Osservatorio Astronomico di Cagliari, via della Scienza 5, I-09047 Selargius (CA), Italy}

\author[0000-0002-1169-7486]{Daniela Huppenkothen}
\affiliation{SRON Netherlands Institute for Space Research, Sorbonnelaan 2, 3584 CA, Utrecht, Netherlands}

\author[0000-0001-5799-9714]{Scott M. Ransom}
\affiliation{National Radio Astronomy Observatory, 520 Edgemont Road, Charlottesville, VA 22903, USA}

\author[0000-0003-1661-0877]{Stefano Curatti}
\affiliation{Infora Societ\`a Cooperativa, Viale Elmas 142, I-09122 Cagliari, Italy}

\author[0000-0001-6762-2638]{Alessandro Ridolfi}
\affiliation{INAF-Osservatorio Astronomico di Cagliari, via della Scienza 5, I-09047 Selargius (CA), Italy}
\affiliation{Max-Planck-Institut f\"{u}r Radioastronomie, Auf dem H\"{u}gel 69, D-53121 Bonn, Germany}

\begin{abstract}

The search for astronomical pulsed signals within noisy data, in the radio band, is usually performed through an initial Fourier analysis to find ``candidate'' frequencies and then refined through the folding of the time series using trial frequencies close to the candidate. 
In order to establish the significance of the pulsed profiles found at these trial frequencies, pulsed profiles are evaluated with a chi-squared test, to establish how much they depart from a null hypothesis where the signal is consistent with a flat distribution of noisy measurements. 
In high-energy astronomy, the chi-squared statistic has widely been replaced by the \zsq statistic and the H-test as they are more sensitive to extra information such as the harmonic content of the pulsed profile. 
The \zsq statistic and H-test were originally developed for the use with ``event data'', composed of arrival times of single photons, leaving it unclear how these methods could be used in radio astronomy.
In this paper, we present a version of the \zsq statistic and H-test for pulse profiles with Gaussian uncertainties, appropriate for radio or even optical pulse profiles.
We show how these statistical indicators provide better sensitivity to low-significance pulsar candidates with respect to the usual chi-squared method, and a straightforward way to discriminate between pulse profile shapes. 
Moreover, they provide an additional tool for Radio Frequency Interference (RFI) rejection.
\end{abstract}


\section{Introduction} \label{sec:intro}
In a typical pulsar search, the uncertainty in the flux measurements from a radio telescope are dominated by various sources of noise, both from the sky and the instruments.
The pulsar signal can be millions of times weaker than the intrinsic noise of the time series. 
Therefore, pulsar searches employ Fourier analysis to search for periodicities.
This search produces a growing number of candidate pulsations (mainly due to the ever increasing interfering signals from Earth or satellites). 
The selection of the most promising astrophysical signals amongst them is based on a small number of statistical indicators. 
Once a candidate pulsar frequency is found, the analysis of candidates typically starts from epoch Folding (EF; \citealt{leahySearchesPulsedEmission1983}).

Let $(t_j, X_j)\, \forall\, j=1, \dots, N$ be pairs of flux measurements $X_j$ at times $t_j$. If we define $f$ as the candidate pulse frequency and $\dot{f}$, $\ddot{f}$, ... the frequency derivatives measured at a reference time $t_{\rm ref}$, we calculate the \textit{pulse phase} of each $X_j$ as

\begin{equation}\label{eq:phase}
    \phi_j(t) = \phi_0 + f (t_j - t_{\rm ref}) + 0.5 \dot{f} (t_j - t_{\rm ref})^2 + ...
\end{equation}

where $\phi_0$ is the pulse phase at $t_{\rm ref}$, which is set to 0 for unknown or candidate pulsars.
Given the periodicity involved, one is typically only interested in the fractional part of this phase, distributed between 0 and 1. 
Since this quantity will always be multiplied by a $2\pi$ factor in the formulae below, we will follow the convention of including $2\pi$ in $\phi$ for consistency: hereafter,
$\phi$ will always mean an angle between 0 and $2\pi$.
The folded or pulsed profile is then a histogram of these phases falling into \nbin equal phase bins between 0 and $2\pi$, weighted by the flux in each sample\footnote{Whereas this is the most common way to fold radio data, some software packages like PRESTO use a slightly different approach: they assume that each sample is finite in duration and "drizzle" it over the appropriate pulse phase bins.
This slightly different approach does not substantively affect the results of this paper, but see Appendix~\ref{app:drizzle} for more details.}:

\begin{equation}
p_i = \sum_{j=1}^N X_j \theta\left(\phi_j - \phi_{{\rm mid}, i}\right) (i = 1, \dots, \nbin)
\label{eq:folding}
\end{equation}

where 
$$\phi_{{\rm mid}, i} = 2\pi \frac{i - 0.5}{\nbin}$$ 
is the phase corresponding to the middle of bin $i$ and

\begin{equation}
    \theta(x) = 
        \begin{cases}
            1 & \mbox{if } |x| < 0.5/\nbin \\
            0 & \mbox{otherwise } 
        \end{cases}
\end{equation}
Given a pulsed profile  $p_i$, consisting of \nbin equally spaced bins, the most common statistical indicator is the Epoch Folding chi-squared statistic \citep{leahy_searches_1983,leahy_searches_1987}

\begin{equation}\label{eq:ef}
    \ef = \sum_{i=1}^{\nbin} {\left(\frac{p_i - \bar{p}}{\sigma_i}\right)}^2
\end{equation}
where $\bar{p}$ is the average profile, and $\sigma_i$ is the standard deviation in each profile bin.
In the typical case, the uncertainties $\sigma_i$ are derived from the statistical uncertainties of the individual flux measurements $X_j$. We assume they are distributed according to the Poisson or normal distribution, that the central limit theorem holds and, therefore, that the phase-folded flux measurements $p_i$ are distributed normally. 
Also, when comparing the pulse profile obtained folding at different periods, one usually assumes that the $\sigma_i$ are all equal and obtained by error propagation from the standard deviation of the $X_j$:

\begin{equation}\label{eq:std}
    \sigma_i \equiv \sigma = \mathrm{std}(X_j) \sqrt{\left(\frac{N_{\rm samples}}{\nbin}\right)}
\end{equation}

If the normality assumption holds, the quantity \ef follows a $\chi^2_{\nbin - 1}$ distribution if the time series is composed of pure noise, and that makes it easy to evaluate the probability of rejecting the null hypothesis.
Another quick statistical indicator, also used in searches of non-periodic impulsive phenomena in the time domain (i.e. Fast Radio Bursts and single pulses), is the simple signal to noise ratio (SNR):

\begin{equation}\label{eq:snr}
    \mathrm{SNR} = \frac{\mathrm{max}(p_i) - \bar{p}}{\sigma}
\end{equation}
where $\sigma$ is the standard deviation or the median absolute deviation of the profile, and the mean can be substituted by the median depending on the implementation.

In the X-ray and $\gamma$-ray bands astronomers have developed other statistical indicators that are more sensitive --better at finding low-amplitude signals-- than \ef, provided that the signal is composed of single ``events'' and the pulsed component can be described as the sum of a relatively small number of sinusoidal harmonics.
Moreover, while \ef and SNR only measure any deviation of the pulsed profile from a flat distribution, these new statistical indicators also provide the ability to discriminate between different pulse shapes, as we will discuss below.

Let us assume that we are observing a $\gamma$-ray or X-ray pulsar we already know. 
We know its ephemeris (i.e., how its rotation evolves over time), usually expressed as an initial phase $\phi_0$ and series of frequency derivatives ($f,\,\dot{f},\,\ddot{f}, ...$) measured at a reference time $t_{\rm ref}$.
Starting from a list of ``events'' corresponding to time stamps $t_j$ of photons hitting a detector, one can calculate the phase of each photon $\phi_j$ as in \eqref{eq:phase}.
The pulsed profile in this case can be obtained as a histogram of the fractional part of these phases.
$p_i$ is now the number of events counted in the phase bin $i$.
\eqref{eq:ef} in this case becomes

\begin{equation}\label{eq:efbin}
    \efbin = \sum_{i=1}^{\nbin} \frac{{\left(p_i - \lambda\right)}^2}{\lambda}
\end{equation}
where $\lambda$ is the average of the pulsed profile (the equivalent of $\bar{p}$ in Equation \ref{eq:ef}), and we used the fact that in a counting experiment the variance in each bin is equal to the Poisson rate.

\citet{buccheri_search_1983} introduced an  alternative statistic, now-standard in X- and $\gamma$-rays, called \zsq, which is generally more sensitive for pulsed signals whose pulsed profile can be described by a small number of sinusoidal harmonics\footnote{Note that this is close but not equivalent to the incoherent harmonic summing used in radio searches, \citep[See, e.g.][]{ransomFourierTechniquesVery2002,lorimerkramer}}.
If we detect $N$ photons and their pulse phase is $\phi_i$ ($i=1..N$), we define \zsq as

\begin{equation}\label{eq:zn}
\zsq =\dfrac{2}{N} \sum_{k=1}^n \left[{\left(\sum_{j=1}^N \cos k \phi_j\right)}^2 + {\left(\sum_{j=1}^N \sin k \phi_j\right)}^2\right]
\end{equation}
This formula evaluates how well the distribution of pulse phases is described by a series of $n$ sinusoidal harmonics of increasing order.
The statistic in the case of a single harmonic, $Z^2_1$, is also known as the Rayleigh test \citep{mardiaStatisticsDirectionalData1975,gibsonTransientEmissionUltrahigh1982}.
Similarly to \ef, this statistic can be used for pulsar searches as well. 
Given a certain number of trial pulse periods, one can calculate \zsq and select the periods that give the highest value for the statistic.
\zsq follows a \chizsq distribution for noise powers (i.e. the statistic's values obtained with trial period values far from the real period), so that it is straightforward to set ``detection'' levels or calculate the probability of rejecting the null hypothesis given a \zsq value.

To show why the \zsq statistic is more sensitive than \efbin for a small number of sinusoidal components, we can follow \citet{leahy_searches_1983} and define the ``quality factor'' of the statistic as

\begin{equation}\label{eq:quality}
Q = \frac{\ssig}{\sthr- \dof}
\end{equation}
where \ssig is the statistic's value of a given pulsed profile, \sthr is the ``threshold'' statistic value indicating a given small p-value for noise powers, and \dof is the number of degrees of freedom. 
\citet{leahy_searches_1983} compares Epoch Folding and the Rayleigh test ($Z^2_1$) using square profiles with different duty cycles (i.e. duration of the ``high'' signal value), showing that \ef is more sensitive for profiles with shorter duty cycle (duration), while the Rayleigh test is more sensitive for profiles with longer duty cycle. 
In the next section, we will use a similar approach to show that \zsq is more sensitive than \ef in all cases where the signal can be described by a relatively small number of harmonics.

Finally, one can, in a single test, determine if the signal is described by a sum of harmonics, and what is the best number of harmonics to describe the signal.
This is known as the $H$ test, and it is basically a summary of multiple \zsq values, properly normalized so that they can be compared \citep{de_jager_poweful_1989,dejagerHtestProbabilityDistribution2010}:

\begin{equation}\label{eq:H}
H = \mathrm{max (Z^2_m - 4m + 4)},\,m=1,20
\end{equation}
with the $m$ corresponding to the maximum usually indicated with $M$.

While these statistical indicators were developed for Poisson-distributed data (photon-counting experiments), they can be adapted to the case of pulsed profiles obtained by, for example, radio telescopes, through a few simple prescriptions.
This paper explains how to adapt these tests to generic pulsed profiles.
We do this in two steps: 1) in \secref{sec:binned} we test that the binned version of the \zsq statistic proposed by \citet{huppenkothenStingrayModernPython2019} has the required statistical properties both with white noise and with pulsed signals; 2) in \secref{sec:gaussians_are_magic} we extend this formulation by introducing a version of \zsq for pulsed profiles with Gaussian uncertainties.

\section{Extension of the Z statistic for binned Poisson data}\label{sec:binned}
\begin{figure}
    \centering
    \includegraphics{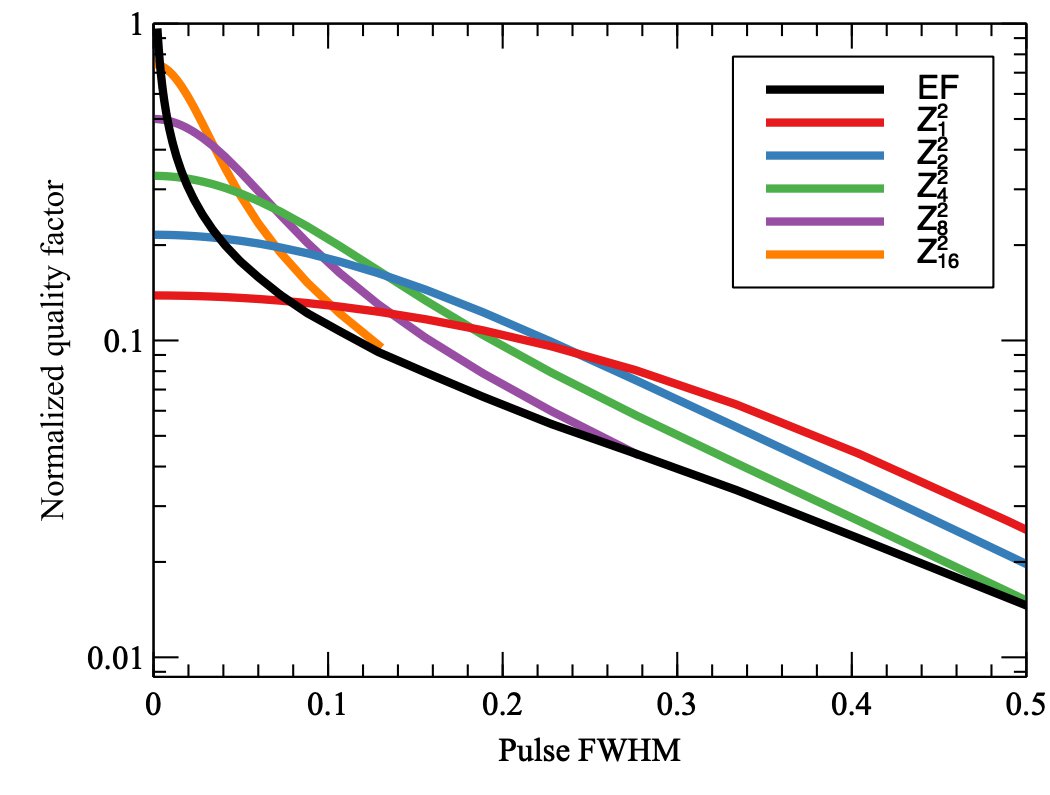}
    \caption{Quality factors for \ef and \zsqbin for binned pulsed profiles consisting of a single Gaussian with increasing width (in units of pulse phase, from 0 to 1). Following \citep{leahy_searches_1983}, we normalized the quality factor by $Nf^2$, where $f$ is the ratio between the pulsed area and the total area under the pulsed profile including the DC level.}
    \label{fig:quality}
\end{figure}

As we have seen, \eqref{eq:zn} defines the \zsq statistic for unbinned event data.
\citealt{huppenkothenStingrayModernPython2019} introduced a version of the \zsq test for binned pulsed profiles obtained from X-ray event data):

\begin{equation}\label{eq:znbin}
\zsqbin \approx \dfrac{2}{\sum_i{p_i}} \sum_{k=1}^n \left[{\left(\sum_{i=1}^{\nbin} p_i \cos k \phi_i\right)}^2 + {\left(\sum_{i=1}^{\nbin} p_i \sin k \phi_i\right)}^2\right]
\end{equation}
where $p_i$, working as a weight, corresponds to the number of photons in a given profile bin.
We know that we need at least two bins for each cycle to sample a sinusoid (Nyquist limit), so we expect the approximation to make little sense if $\nbin < 2n$.
Also, it is reasonable to expect a better approximation as we sample each harmonic cycle with more and more bins.
\citet{huppenkothenStingrayModernPython2019} prudently recommend to bin the pulsed profile with a number of bins at least 10 times larger than the number of harmonics $n$, but give no justification.

The effect of binning can be quantified through simple analytical manipulations (developed in Appendix~\ref{sec:difference}).
Calling 
\begin{eqnarray}
A_k = \sum_{j=1}^{N} \cos k\phi_j\\
B_k = \sum_{j=1}^{N} \sin k\phi_j
\end{eqnarray}
we can express \zsq as 

\begin{equation}
\zsq =\dfrac{2}{N} \sum_{k=1}^n \left[A_k^2 + B_k^2\right]
\end{equation}

One can demonstrate (see Appendix~\ref{sec:difference}) that the binned approximation gives

\begin{equation}\label{eq:binapprox}
    \zsqbin \approx 
    \frac{2}{N}\sum_{k=1}^n {\left(\frac{\nbin}{\pi k} \sin \frac{\pi k}{\nbin} \right)}^2
    \left[A_k^2 + B_k^2\right]
\end{equation}
Comparing it to \eqref{eq:znbin}, we see that this formula only differs by the factors
\begin{eqnarray}
    C(k) &=& {\left(\frac{\nbin}{\pi k} \sin \frac{\pi k}{\nbin} \right)}^2\label{eq:correction}
\end{eqnarray}
that go to 1 for $k\ll \nbin$ and lead to a loss of sensitivity\footnote{This is equivalent to the loss of sensitivity in the power density spectrum due to sampling. 
See \citet{vanderklisFourierTechniquesXray1989}, Appendix~\ref{sec:difference}} of $\sim40\%$ for $k=\nbin/2$.


Assuming a pulsed profile of the kind 

\begin{equation}
    p_i = \lambda \left[1 + \sum_{l=1}^m a_l \sin l(\phi_i - \phi_{0, l})\right]
\end{equation}
(i.e. a DC level $\lambda$ plus a finite number $m$ of sinusoidal harmonics with random phases) we can define a statistic \ssig as

\begin{equation}\label{eq:ssig}
\ssig \approx \frac{\lambda\nbin}{2}\sum_{l=1}^{m} a_l^2 \; .
\end{equation}
It can be shown (See Appendices~\ref{sec:sumzsq} and~\ref{sec:sumef}) that the statistic \ssig has exactly the same value for \ef and \zsqbin provided that $n\geq m$ and one can neglect the binning ($\nbin \gg 2m$) and reduces to the $(1/2) N A^2$ (where $N=\lambda \nbin$ is the total number of photons) given by \citet{leahy_searches_1983}, for $m=1$.

Taking \eqref{eq:quality} and comparing the sensitivity of \ef and \zsqbin, we find the results plotted in \figref{fig:quality}.
Here we created Gaussian pulsed profiles with increasing width (and so, a decreasing number of harmonics) and calculated the following statistic:
\begin{itemize}
    \item \zsqbin for n=1, ... 20
    \item \ef using an increasing number of bins to avoid the sensitivity loss when $\nbin/n \sim2$
\end{itemize}
We used a number of bins that oversampled the Gaussian peak (two points inside $1-\sigma$) in order not to decrease the sensitivity of any of the methods.
Therefore, in \figref{fig:quality}, \nbin increases going from right to left.
As expected, \zsqbin is \textit{always} more sensitive than \ef when we use a sufficient number of sinusoidal components.
The reason is that the \zsqbin threshold level for any given $n$ is insensitive to the number of bins, if not for small corrections at small \nbin, while \ef's threshold increases with \nbin. 
To detect very sharp signals (or high-order harmonics) we need a large number of profile bins, and \ef is systematically noisier than \zsqbin.
Since the signal level \ssig is the same for \zsqbin and \ef (\eqref{eq:ssig}), increasing quadratically with signal amplitude, and the denominator increases slowly with increasing number of degrees of freedom (approximately with the square root), the significance of very strong signals will be very similar using the two methods.

\zsqbin can be used to calculate the H statistic, a single test which allows to investigate the occurrence of signals of multiple sinusoids whose number is not known a priori (\eqref{eq:H}).
Our generalization of the \zsq test to binned data substantially confirms what was described by \citet{de_jager_poweful_1989} for the unbinned statistic, and validates the use of the H test for binned profiles.
In the next section, we further extend this method to generic profiles (i.e. obtained by any measurements and not just particle/photon counts), provided that their measurement error is approximately Gaussian and constant throughout the profile.
Hereafter, we will run the \zsqbin analysis using multiple $n$, and we will be using the $H$-test to decide for each profile what number of harmonics $M$ gives the best detection.
Therefore, when we talk about the \zsqbin significance or the $H$-test significance, we will be referring to the significance of \zsqbin with $n=M$.

\section{The Z statistic in the case for normally distributed data}
\label{sec:gaussians_are_magic}

Let us start with a set of $N$ observations $\{ X_j \}_j^N$ at times $t_j$, as defined in Section \ref{sec:intro}, distributed around a mean flux $\mu_x$ with a variance of $\sigma^2_x$. We assume here that the variance is the same for all bins, $\sigma^2_{x,j} = \sigma^2_x$, but will relax that requirement later. We aim to use the $Z^2$ statistic to compare to a null hypothesis of a constant flux, hence we also assume that the measurements are distributed around a constant flux value $\mu_x = \mathrm{constant}$. Thus, measurements are drawn from a normal distribution, $x_j \sim \mathcal{N}(\mu_x, \sigma_x^2)$. 

We now compute a pulse profile by distributing fluxes into \nbin phase bins, and sum the fluxes within each phase bin to obtain phase-dependent fluxes 

\begin{equation}
p_i = \sum_{j=i\nfmpb}^{(i+1)\nfmpb}{x_j} \; \forall \; i, \, i=0,\dots,\nbin-1\, . 
\end{equation}

Here, $\nfmpb$ is the number of flux measurements in each phase bin, $\nfmpb = \frac{N}{\nbin}$. Because the sum of Gaussian variables is also distributed as a Gaussian, $p_i$, too, are distributed normally, with the same mean $\mu_p = \sum_{j=1}^{\nfmpb}{\mu_x} =  \mu_x \nfmpb$. The variance of the summed random variable can also be calculated as the sum of variances:

\[
\sigma^2_p = \sum_{j=1}^{\nfmpb}{\sigma_x^2} = \sigma_x^2 \nfmpb  \; .
\]

\noindent We define the \zsqgauss statistic as 

\begin{equation}
\zsqgauss^{2} = \frac{1}{K} \sum_{k=1}^{n}\left[ \left( \sum_{i=1}^{\nbin} {p_i \cos{k\phi_i}}  \right)^2 + \left( \sum_{i=1}^{\nbin} {p_i \sin{k\phi_i}}\right)^2\right]
\label{eq:zngauss}
\end{equation}

\noindent \added{where $K = \sigma_P^2 \nbin/2$ (see below),}
$p_i$ are phase-folded fluxes in each of \nbin phase bins. Each phase bin is characterized by a phase $\phi_i$ measured at the mid-point of the phase bin. Finally, in this general case, we compute the $Z^2$ statistic over $n$ harmonics.
In the following, we will assume $n=1$ for much of this section, but will show in the end how this case generalizes to $n>1$.

In the unbinned Poisson case for which $Z^2_n$ was initially defined, the phases $\phi_j$ are uniform random variables across the interval $\left[0, 2\pi \right]$. For the case of summed Gaussian fluxes, the phases $\phi_i$ are not random variables, but rather real numbers on a regular grid over the same interval spanning $0$ to $2\pi$. It follows that the sine and cosine of the phases $\sin{\phi_i}$ and $\cos{\phi_i}$ are also not random variables, but real numbers over the interval $\left[-1, 1\right]$. In the following, we will first derive the appropriate distributions for the cosine term, but because trigonometric identities hold, the sine term will produce random variables drawn from the same statistical distribution. 

Because $\cos{\phi_i}$ is not a random variable and $p_i$ is Gaussian, the product will also be distributed normally, with 

\[
p_i \cos{\phi_i} \sim \mathcal{N}(\mu_p \cos{\phi_i}, \sigma_P^2 \cos^2{\phi_i})
\]

The sum of the product of summed fluxes $p_i$ and cosine terms amounts to another sum of random variables drawn from a number of independent and identically distributed Gaussians, each of which has been scaled by $\cos{\phi_i}$. We define the sum as

\[
A = \sum_{i=1}^{\nbin} {p_i \cos{k\phi_i}}
\]

\noindent This sum, too, yields a Gaussian distribution for which mean and variance are defined as

\begin{eqnarray}
\label{eq:sum_moments}
\mu_\mathrm{A} & = &  \sum_{i=1}^{\nbin}{\mu_p \cos{\phi_i}} \\ \nonumber
\sigma^2_\mathrm{A} & = & \sum_{i=1}^{\nbin}{\sigma^2_p \cos^2{(\phi_i)}} \; .
\end{eqnarray}
 
 \noindent Note that because $\sum_{i=1}^{\nbin}{\cos{\phi_i}} = 0$, it follows that $\mu_\mathrm{A} = 0$. Similarly, for a variance that is constant across all profile bins, we can take $\sigma^2_p$ outside of the sum, and calculate the value for $\sum_{i=1}^{\nbin}{\cos^2{(\phi_i)}}$ as a Riemann sum: 
 
 \[
     \sum_{i=1}^{\nbin}{\cos^2{(\phi_i)}} = \frac{\pi}{\Delta \phi}
 \]

\noindent where $\Delta\phi$ is the size of a phase bin, $\Delta \phi = \frac{2\pi}{\nbin}$. Thus, we find that 

\begin{equation}
\label{eq:sum_variance}
\sigma^2_\mathrm{A} = \sigma_P^2 \frac{\pi \nbin}{2 \pi} = \sigma_P^2 \frac{\nbin}{2}
\end{equation}

\noindent Based on this result, we can now determine the statistical distribution for $A^2$. In particular, for a Gaussian random variable $X$ with a standard distribution $\sigma$, the variable normalized by its standard deviation is distributed following a standard chi-square distribution with one degree of freedom: $\frac{X^2}{\sigma^2} \sim \chi^2_1$. It thus follows that we can define a variable based on $A^2$, where $A$ is defined above, that follows a standard $\chi^2_1$ distribution.

If

\[
A^2 = \left(\sum_{i=1}^{\nbin} {p_i \cos{k\phi_i}}\right)^2
\]

\noindent then the variable

\begin{equation}
Q = \frac{A^2}{\sigma^2_\mathrm{A}} = \frac{\left(\sum_{i=1}^{\nbin} {p_i \cos{k\phi_i}}\right)^2}{\sigma_P^2 \frac{\nbin}{2}}
\end{equation}

\noindent follows a standard $\chi^2_1$ distribution. We can treat the second term in the outer sum of the $Z^2$ statistic, 

\[
B = \sum_{i=1}^{\nbin} {p_i \sin{k\phi_i}}
\]

\noindent in the same way as $A$. Because trigonometric identities hold, we obtain another $\chi^2_1$ distributed variable for $B^2$. 

Adding these (properly normalized) variables together yields the final $Z^2$ statistic. Because the sum of $\chi^2_1$ distributed variables is a $\chi^2_2$ distributed variable with degrees of freedom $df = 2$, we find that $Z^2_1 \sim \chi^2_2$ assuming the correct normalization constant $K$, defined for the case of constant data variances, $\sigma_{P,j} = \sigma_P$ as

\begin{equation}
K = \sigma_P^2 \frac{\nbin}{2}
\label{eq:normalization}
\end{equation}

\noindent Extending this result to $n>1$ in order to account for multiple harmonics is straightforward: each harmonic adds a variable to the sum drawn from $\chi^2_2$. Because the sum of $k$ random variables, each drawn from a $\chi^2_2$ distribution, is another $\chi^2$ distribution with $df = 2n$, we find that

\begin{equation}
    \zsqgauss \sim \chi^2_{2n} \; .
\end{equation}

\subsection{Extension to data sets where the uncertainty varies with phase}
\label{sec:phaseuncertainty}

Let us assume a case where $\sigma^2_{p} = \sigma^2_{p,i}$, i.e. the summed fluxes are still drawn from a normal distribution, but each with a different variance, 

\[ 
p_i \sim \mathcal{N}(\mu_x, \sigma_{p,i}^2) \; .
\]

\noindent This change in assumptions alters one important step of the calculation: in \eqref{eq:sum_variance}, we have assumed that the variance of each profile bin is identical, which made it possible to take that term out of the sum and compute the sum of all $\cos{\phi_i}$ terms independently. For unequal variances, this no longer holds, and normalization for the $Z^2_n$ statistic instead becomes

\begin{equation}
   K = \sum_{i=1}^{\nbin}{\sigma_{p,i}^2 \; . \cos^2(\phi_i)}
\end{equation}

\noindent Note that this normalization is a generalization of \eqref{eq:normalization}.

The distributions presented above are \textit{exact}. For both unbinned and binned Poisson data, the assumption that we can write $Z^2_n \sim \chi^2_{2n}$ crucially rests on the validity of the Central Limit Theorem for this case. In the case of unbinned Poisson data, the phases $\phi_j$ are uniformly distributed random variables, and the cosine of these phases is distributed as 
\[
\cos{\phi_j} \sim \mathrm{Arcsin}(-1, 1)
\]

\noindent where $\mathrm{Arcsin}(-1, 1)$ is a scaled arcsine distribution (and equivalently, $\sin{\phi_j}$ also produces an arcsine distribution). This distribution is heavily non-Gaussian, and the sum $\sum_{j=1}^{N}{\cos{\phi_j}}$ for $N$ photons will only approximate a Gaussian distribution if $N$ is reasonably large.

Similarly, when binned Poisson event rates are used in \eqref{eq:zngauss} as $p_i$ instead of normally distributed measurements, the resulting product $p_i \cos{\phi_i}$ will not be drawn from any analytically known probability distribution. As a result, $Z^2_{n}$ will only be distributed as the expected $\chi^2_{2n}$ distribution if either the count rates $p_i$ in each bin are large enough that the Poisson counts approximate a normal distribution, or there are a sufficient number of bins $i$ that the sum of bins will again tend to a Gaussian.

Also note that in the binned Poisson case with mean and variance equal to $\lambda$, \eqref{eq:normalization} yields

\[
K = \sigma_p^2 \frac{\nbin}{2} = \frac{\lambda\nbin}{2} = \frac{\sum_i p_i}{2}
\]
that confirms \eqref{eq:znbin}.

\subsection{Extension to unevenly sampled Gaussian-distributed measurements}

In practice, light curves--especially in the optical wavelength regime--are often subject to observing constraints, leading to unevenly sampled time series. In calculating $Z^2$, this translates into phase-binned fluxes $p_i$ for which the number of measurements in each bin differs for each bin $i$, $\nfmpb = \nfmpbi$. In these cases, it is advisable to calculate $Z^2$ for the \textit{average} of fluxes falling into the same phase bin, rather than the \textit{sum}, as we have done above. 

In this case, we define

\begin{equation}
    p_i = \frac{1}{\nfmpbi} \sum_{j=1}^{\nfmpbi}{x_j} \; \forall \; i, \, i=0,\dots,\nbin-1\, 
\end{equation}

In this case, the mean of the resulting variables $p_i$ correspond to the mean of the individual measurements, $\mu_p = \mu_{x}$, and the variance becomes

\begin{equation}
    \sigma^2_{p,i} = \frac{1}{\nfmpbi}\sigma^2_x \; .
\end{equation}

\noindent The remainder of the procedure can be calculated using the same formalism as for the case described in Section \ref{sec:phaseuncertainty}.

\section{Test the Gaussian Z and H tests: simulations}\label{sec:norm}

Let us now test whether this method produces predictable values of \zsq and H, and compare them with the other statistical indicators used in radio astronomy.




We simulated 10000 pulsed profiles with 1024 bins with Gaussian errors, varying all parameters (input SNR, peak width, baseline, noise, etc.). 
The large number of bins was chosen to minimize binning effects for high-order harmonics (See \secref{sec:binned}, \eqref{eq:binapprox}.
We evaluated the profiles with the following statistic:
\begin{itemize}
    \item the Epoch folding statistic (\eqref{eq:ef}), rebinning the pulsed profile to 8, 16, 32, ... 256 bins. 
    As the EF statistic is noisier for a higher number of bins, but on the other hand it might miss sharp profiles for a small number of bins, we tried various combinations to be sure we were not underplaying this technique.
    \item the \zsq and H statistics (Equations~\ref{eq:zngauss} and~\ref{eq:H}).
\end{itemize}

\begin{figure}
\centering
\includegraphics[width=\linewidth]{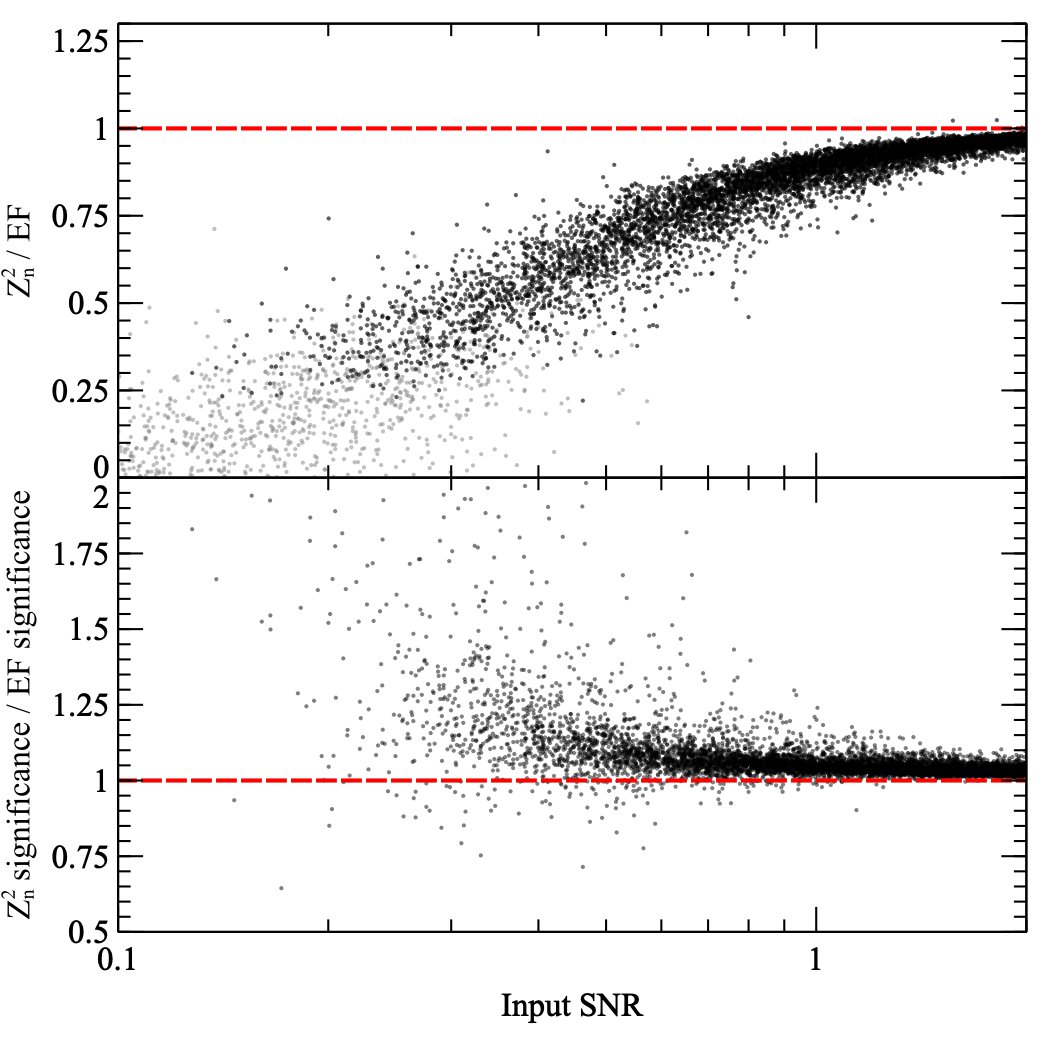}
\caption{Comparison of the raw statistical value (top) and the detection significance (bottom) using the standard epoch folding statistic (\eqref{eq:ef}) and the \zsqgauss (\eqref{eq:zngauss}).
In the signal-dominated regime  (high significance) the values from \ef and \zsqgauss, and their significances, approach asymptotically. 
However, at \textit{low} SNR the \ef gives higher values of the statistic, but \zsq is more sensitive, yielding a higher significance (see \secref{sec:binned}).
Grey points indicate values below detection level (these points have invalid significance ratio and are absent in the bottom plot.)
\label{fig:detection}}
\end{figure}
The results are plotted in \figref{fig:detection}.
As the plot shows, the \zsq statistic has equivalent sensitivity to the EF statistic in the signal-dominated regime, with the advantage that it allows to classify the signals based on their harmonic content. 
However, the \zsq is \textit{more} sensitive at low significance.
This is because the noise spectrum of \zsq is less noisy (For $2n<n_{bin}$, which is always the case), and a more or less equivalent signal power will ``stand out'' more clearly against the noise.

By using the H test, we can run the \zsq for multiple values of $n$ and obtain the best number of harmonics to describe the profile.

Thanks to these tests, we are confident that the method described in this Section, with the formulation described in \secref{sec:binned}, allows one to apply the \zsq and H statistics to pulsar searches conducted with a non-counting instrument producing data with normally distributed uncertainties, for example in the radio or optical band.

\section{Applying the method to real data}

The final and most important test is applying the methods described in \secref{sec:gaussians_are_magic} to real astronomical data.
For this, we used data from a test observation of PSR\ B0331+45 taken at the Sardinia Radio Telescope (SRT; \citealt{prandoniSardiniaRadioTelescope2017}). 
PSR\ B0331+45 \citep{deweySearchLowluminosityPulsars1985} is a relatively slow pulsar ($p_{\rm spin}\sim269$\, ms), with a very sharp profile.
We selected a 30-min observation at L-band (1300-1800 MHz) performed on UT 2016-01-09 as part of routine calibration procedures for a pulsar survey. The data were acquired using the pulsar DFB3 backend\footnote{\url{www.jb.man.ac.uk/pulsar/observing/DFB.pdf}}) with a frequency resolution of 2 MHz and a time resolution of 128 ms. 

\subsection{Data processing and pulsar search}\label{sec:prestosearch}
We analyzed the data with \texttt{PULSAR\_MINER}\footnote{\url{https://github.com/alex88ridolfi/PULSAR_MINER}}, a pipeline based on PRESTO \citep{ransomPRESTOPulsaRExploration2011,ransomFourierTechniquesVery2002}.
The observation was plagued by strong RFI, both impulsive and periodic, broad- and narrow-band, which could not be eliminated completely using PRESTO's tool \texttt{rfifind}.
A standard blind search was performed, under the hypothesis that we did not know anything about the pulsar we are looking for. The data were dedispersed with 980 dispersion measure (DM) trials, in the range 2 to 100 pc\,cm$^{-3}$. We then used \texttt{accelsearch} to search for periodicities from 1 ms to 20 s within each dedispersed time series. The candidate periods were sifted in order to purge them from the most prominent false positives. 
The final candidates from the search are also saved as text files (with extension \textit{.bestprof}), containing the pulse profile and information on the detection including the period and period derivatives, the reference time, the detection significance, an estimate of the noise of the profile, etc.
Finally, we applied the methods described in \secref{sec:gaussians_are_magic} to the profiles of the candidates obtained from this search.

\subsection{Estimating the noise}
Arguably the most important part of applying the method is an estimate of the profile noise, $\sigma^2_p$.
The standard deviation of the profile includes contributions from both the random noise of the data and the signal variability. 
Separating these two components from a folded profile is in general complicated and requires subtraction of a model of the profile.
However, PRESTO correctly calculates the noise a priori, from the variability of the data before folding. 
We recommend that people using different software, in particular custom-made software, do a similar procedure to calculate the standard deviation from \eqref{eq:std} or, equivalently, the profile variance $\mathrm{VAR}(p)$ starting from the data variance $\mathrm{VAR}(d)$:
\begin{equation}
    \mathrm{VAR}(p) =  \mathrm{VAR}(d) \frac{N_{\rm samples}}{\nbin}
\end{equation}

where $N_{\rm samples}$ is the number of samples in the light curve.

\subsection{Analysis of search candidates}

\begin{figure}
\centering
\includegraphics[width=\linewidth]{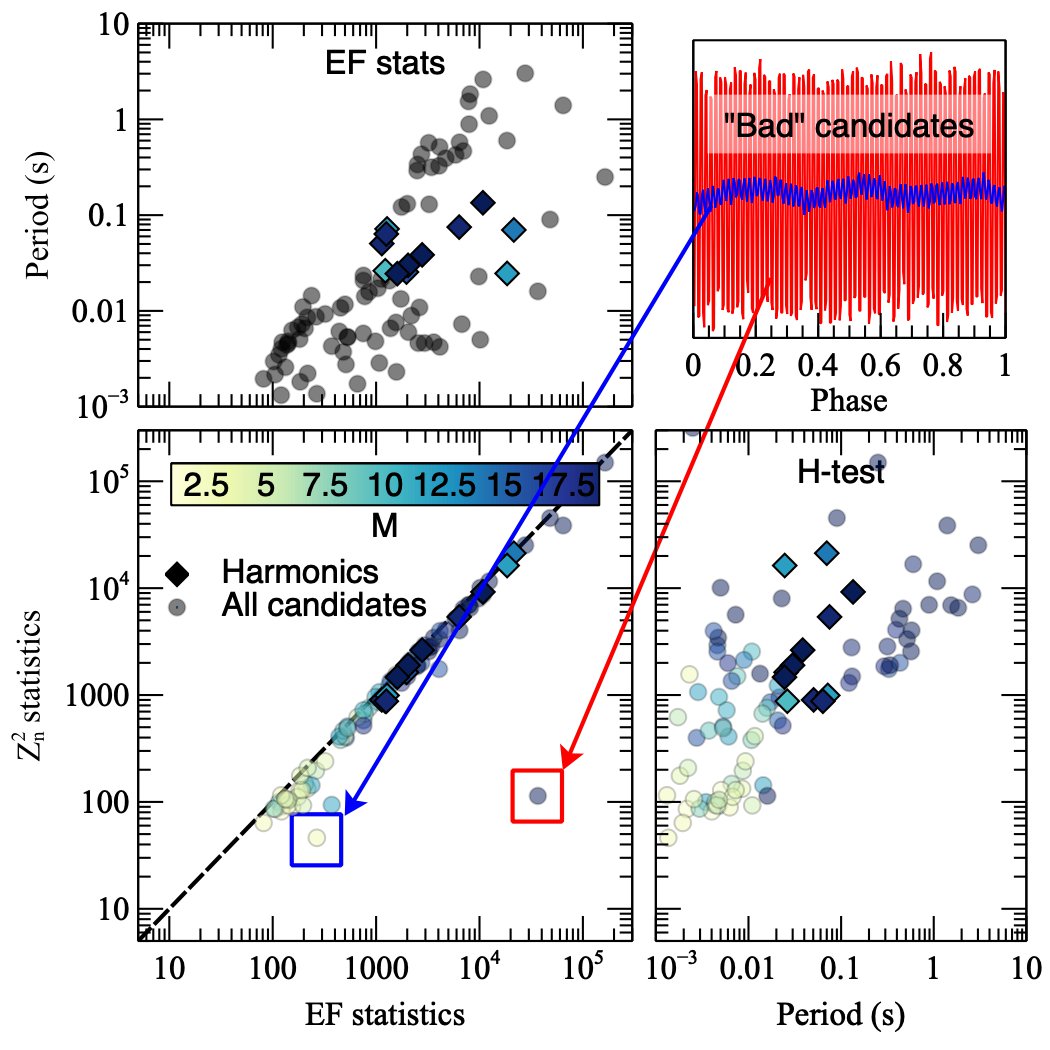}
\caption{Raw values of the statistics from \zsqgauss and \ef for a pulsar search at the Sardinia Radio Telescope.
Colors indicate the number of harmonics corresponding to the best detection in the \zsq search, and
diamonds indicate the harmonics of pulsar candidates (we maintained the color coding of these candidates in the EF panel for an easier comparison).
The \ef statistic shows significantly higher values of significance only for signals with highly distorted pulse shapes, as expected, due to the high number of harmonics needed to describe such shapes. Two examples of such signals are shown, folded, in the upper right panel (note that both pulses have a fast square wave from RFI superimposed to the profile).
\label{fig:compare_sig}}
\end{figure}

\begin{figure}
\centering
\includegraphics[width=\linewidth]{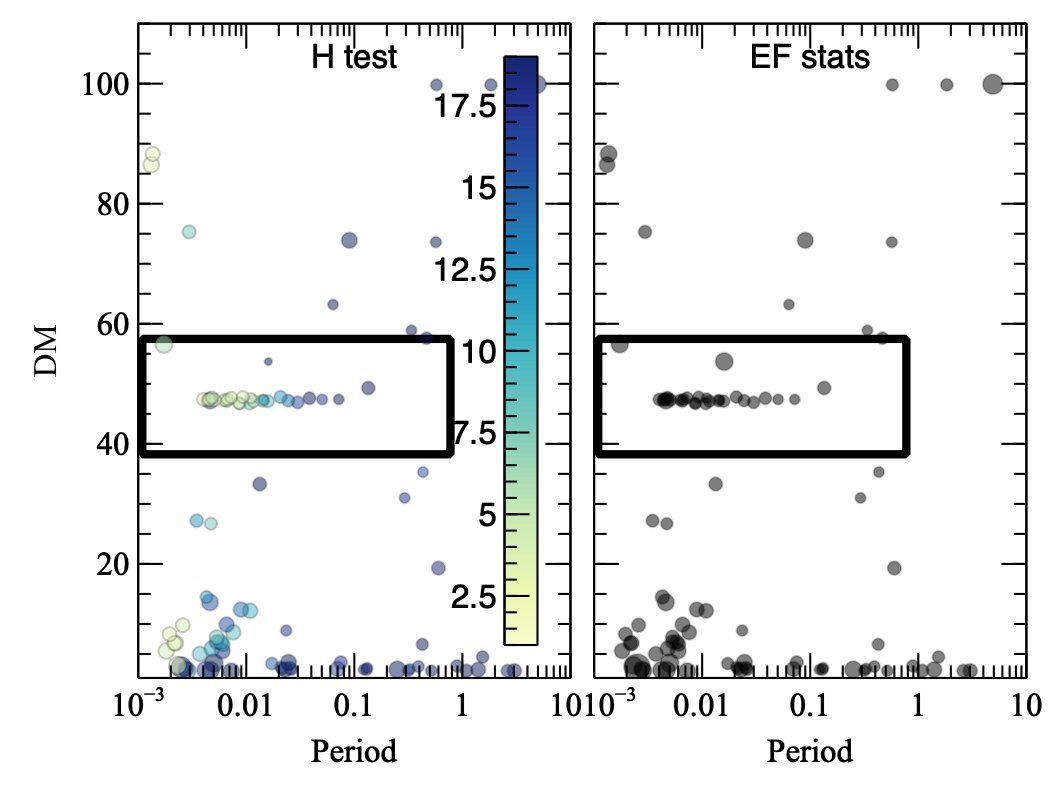}
\caption{Standard DM versus period plot of the candidates of the pulsar search in \figref{fig:compare_sig}, using \ef and \zsqgauss.
The signals being highly significant, the significance with the two methods is mostly equivalent. However, the \zsqgauss allows to easily classify the candidates in terms of shape, and penalizes profiles containing strong square-wave RFI.
\label{fig:real_search}}
\end{figure}

Finally, we applied the statistic to the pulsed profiles obtained in \secref{sec:prestosearch}.
In Figures~\ref{fig:compare_sig} and ~\ref{fig:real_search} we show the results.
As expected, similarly to what was found using simulated data, the values of \zsqgauss and \ef are generally very close for candidates with high SNR. 
We checked the candidates where this was not true and found that, invariably, these candidates had very distorted pulsed profiles, usually with the profile being or containing a square wave (\figref{fig:compare_sig}).
This tells us that a comparison between \ef and \zsqgauss statistics is a good diagnostic to eliminate some of the strong RFI still present among the sifted candidates.
\figref{fig:real_search} shows the candidates in a DM vs period plot, where the color-coding represents the number of harmonics needed to describe the profile, while the size represents the significance of the candidates as calculated using the H test.
This, again, turns out to be a useful way to plot candidates: the candidates corresponding to higher harmonics of the pulsar show up at more and more sinusoidal candidates at the same DM.

\section{Conclusions}
In this paper, we outlined a method to apply statistical tests developed for counting experiments to the folded profiles of radio pulsars.
We did this in two steps: 1) we showed that the \zsq statistic can be easily applied to folded profiles from high-energy pulsars, obtained by counting the events falling at different pulse phases; 2) we applied a normalization to radio pulsar profiles that preserves the signal-to-noise ratio and creates profiles with the correct statistical properties to apply the binned version of \zsq developed at step 1.
We demonstrate how the \zsqgauss statistic can be a great tool to characterize the candidates from radio pulsar searches, being at once more sensitive to low-significance signals and better at discriminating between pulse shapes.
We applied the method to a real pulsar search, showing how the method is also a great tool to eliminate RFI candidates.
\added{The \zsqgauss statistic is available in the open-source software packages for astronomical time series analysis PRESTO \citep{ransomPRESTOPulsaRExploration2011}, Stingray \citep{huppenkothenStingraySpectraltimingSoftware2016,huppenkothenStingrayModernPython2019}, and HENDRICS \citep{bachettiHENDRICSHighENergy2018}.}

\acknowledgments
This work was conducted in the framework of  CICLOPS -- Citizen Computing Pulsar Search, a project supported by \textit{POR FESR Sardegna 2014 – 2020 Asse 1 Azione 1.1.3} (code RICERCA\_1C-181), call for proposal "Aiuti per Progetti di Ricerca e Sviluppo 2017" managed  by Sardegna Ricerche.
The authors wish to thank the referee for useful comments, Paul Demorest and Walter Brisken for the development of PRESTO's effective degree of freedom correction described in Appendix~\ref{app:drizzle}. D.H.  is supported by the Women In Science Excel (WISE) programme of the Netherlands Organisation for Scientific Research (NWO).
AR gratefully acknowledges financial support by the research grant ``iPeska'' (P.I. Andrea Possenti) funded under the INAF national call Prin-SKA/CTA approved with the Presidential Decree 70/2016, and continuing valuable support from the Max-Planck Society.

\software{Stingray \citep{huppenkothenStingraySpectraltimingSoftware2016,huppenkothenStingrayModernPython2019},
          HENDRICS \citep{bachettiHENDRICSHighENergy2018},
          astropy \citep{astropycollaborationAstropyProjectBuilding2018},  
          PINT \citep{luo_pint:_2019},
          ATNF pulsar catalogue \citep{manchesterAustraliaTelescopeNational2005a},
          PRESTO \citep{ransomPRESTOPulsaRExploration2011},
          Veusz\footnote{\url{https://veusz.github.io}}
          }
          

\bibliography{general_z_h}{}

\begin{thebibliography}{}
\expandafter\ifx\csname natexlab\endcsname\relax\def\natexlab#1{#1}\fi
\providecommand{\url}[1]{\href{#1}{#1}}
\providecommand{\dodoi}[1]{doi:~\href{http://doi.org/#1}{\nolinkurl{#1}}}
\providecommand{\doeprint}[1]{\href{http://ascl.net/#1}{\nolinkurl{http://ascl.net/#1}}}
\providecommand{\doarXiv}[1]{\href{https://arxiv.org/abs/#1}{\nolinkurl{https://arxiv.org/abs/#1}}}

\bibitem[{{Astropy Collaboration} {et~al.}(2018){Astropy Collaboration},
  {Price-Whelan}, Sip{\H o}cz, G{\"u}nther, Lim, Crawford, Conseil, Shupe,
  Craig, Dencheva, Ginsburg, VanderPlas, Bradley, {P{\'e}rez-Su{\'a}rez}, {de
  Val-Borro}, Aldcroft, Cruz, Robitaille, Tollerud, Ardelean, Babej, Bach,
  Bachetti, Bakanov, Bamford, Barentsen, Barmby, Baumbach, Berry, Biscani,
  Boquien, Bostroem, Bouma, Brammer, Bray, Breytenbach, Buddelmeijer, Burke,
  Calderone, Cano~Rodr{\'i}guez, Cara, Cardoso, Cheedella, Copin, Corrales,
  Crichton, D'Avella, Deil, Depagne, Dietrich, Donath, Droettboom, Earl, Erben,
  Fabbro, Ferreira, Finethy, Fox, Garrison, Gibbons, Goldstein, Gommers, Greco,
  Greenfield, Groener, Grollier, Hagen, Hirst, Homeier, Horton, Hosseinzadeh,
  Hu, Hunkeler, Ivezi{\'c}, Jain, Jenness, Kanarek, Kendrew, Kern, Kerzendorf,
  Khvalko, King, Kirkby, Kulkarni, Kumar, Lee, Lenz, Littlefair, Ma, Macleod,
  Mastropietro, McCully, Montagnac, Morris, Mueller, Mumford, Muna, Murphy,
  Nelson, Nguyen, Ninan, N{\"o}the, Ogaz, Oh, Parejko, Parley, Pascual, Patil,
  Patil, Plunkett, Prochaska, Rastogi, Reddy~Janga, Sabater, Sakurikar,
  Seifert, Sherbert, {Sherwood-Taylor}, Shih, Sick, Silbiger, Singanamalla,
  Singer, Sladen, Sooley, Sornarajah, Streicher, Teuben, Thomas, Tremblay,
  Turner, Terr{\'o}n, {van Kerkwijk}, {de la Vega}, Watkins, Weaver, Whitmore,
  Woillez, Zabalza, \& {Astropy
  Contributors}}]{astropycollaborationAstropyProjectBuilding2018}
{Astropy Collaboration}, {Price-Whelan}, A.~M., Sip{\H o}cz, B.~M., {et~al.}
  2018, The Astronomical Journal, 156, 123, \dodoi{10.3847/1538-3881/aabc4f}

\bibitem[{Bachetti(2018)}]{bachettiHENDRICSHighENergy2018}
Bachetti, M. 2018, Astrophysics Source Code Library, ascl:1805.019

\bibitem[{Buccheri {et~al.}(1983)Buccheri, Bennett, Bignami, Bloemen,
  Boriakoff, Caraveo, Hermsen, Kanbach, Manchester, Masnou, Mayer-Hasselwander,
  Ozel, Paul, Sacco, Scarsi, \& Strong}]{buccheri_search_1983}
Buccheri, R., Bennett, K., Bignami, G.~F., {et~al.} 1983, A\&A, 128, 245.
\newblock
  \url{http://adsabs.harvard.edu/cgi-bin/nph-data_query?bibcode=1983A%26A...128..245B&link_type=ABSTRACT}

\bibitem[{{de Jager} \&
  B{\"u}sching(2010)}]{dejagerHtestProbabilityDistribution2010}
{de Jager}, O.~C., \& B{\"u}sching, I. 2010, Astronomy and Astrophysics, 517,
  L9, \dodoi{10.1051/0004-6361/201014362}

\bibitem[{de~Jager {et~al.}(1989)de~Jager, Raubenheimer, \&
  Swanepoel}]{de_jager_poweful_1989}
de~Jager, O.~C., Raubenheimer, B.~C., \& Swanepoel, J. W.~H. 1989, 221, 180.
\newblock
  \url{http://adsabs.harvard.edu/cgi-bin/nph-data_query?bibcode=1989A%26A...221..180D&link_type=ABSTRACT}

\bibitem[{Dewey {et~al.}(1985)Dewey, Taylor, Weisberg, \&
  Stokes}]{deweySearchLowluminosityPulsars1985}
Dewey, R.~J., Taylor, J.~H., Weisberg, J.~M., \& Stokes, G.~H. 1985, The
  Astrophysical Journal Letters, 294, L25, \dodoi{10.1086/184502}

\bibitem[{Gibson {et~al.}(1982)Gibson, Harrison, Kirkman, Lotts, Macrae,
  Orford, Turver, \& Walmsley}]{gibsonTransientEmissionUltrahigh1982}
Gibson, A.~I., Harrison, A.~B., Kirkman, I.~W., {et~al.} 1982, Nature, 296,
  833, \dodoi{10.1038/296833a0}

\bibitem[{Huppenkothen {et~al.}(2016)Huppenkothen, Bachetti, Stevens, Migliari,
  \& Balm}]{huppenkothenStingraySpectraltimingSoftware2016}
Huppenkothen, D., Bachetti, M., Stevens, A.~L., Migliari, S., \& Balm, P. 2016,
  Astrophysics Source Code Library, ascl:1608.001

\bibitem[{Huppenkothen {et~al.}(2019)Huppenkothen, Bachetti, Stevens, Migliari,
  Balm, Hammad, Khan, Mishra, Rashid, Sharma, Ribeiro, \&
  Blanco}]{huppenkothenStingrayModernPython2019}
Huppenkothen, D., Bachetti, M., Stevens, A.~L., {et~al.} 2019, ApJ, 881, 39,
  \dodoi{10.3847/1538-4357/ab258d}

\bibitem[{Leahy(1987)}]{leahy_searches_1987}
Leahy, D.~A. 1987, 180, 275.
\newblock
  \url{http://adsabs.harvard.edu/cgi-bin/nph-data_query?bibcode=1987A%26A...180..275L&link_type=ABSTRACT}

\bibitem[{Leahy {et~al.}(1983{\natexlab{a}})Leahy, Darbro, Elsner, Weisskopf,
  Kahn, Sutherland, \& Grindlay}]{leahySearchesPulsedEmission1983}
Leahy, D.~A., Darbro, W., Elsner, R.~F., {et~al.} 1983{\natexlab{a}}, ApJ, 266,
  160, \dodoi{10.1086/160766}

\bibitem[{Leahy {et~al.}(1983{\natexlab{b}})Leahy, Elsner, \&
  Weisskopf}]{leahy_searches_1983}
Leahy, D.~A., Elsner, R.~F., \& Weisskopf, M.~C. 1983{\natexlab{b}}, ApJ, 272,
  256, \dodoi{10.1086/161288}

\bibitem[{{Lorimer} \& {Kramer}(2012)}]{lorimerkramer}
{Lorimer}, D.~R., \& {Kramer}, M. 2012, {Handbook of Pulsar Astronomy}

\bibitem[{Luo {et~al.}(2019)Luo, Ransom, Demorest, van Haasteren, Ray, Stovall,
  Bachetti, Archibald, Kerr, Colen, \& Jenet}]{luo_pint:_2019}
Luo, J., Ransom, S., Demorest, P., {et~al.} 2019, Astrophysics Source Code
  Library, ascl:1902.007.
\newblock \url{http://adsabs.harvard.edu/abs/2019ascl.soft02007L}

\bibitem[{Manchester {et~al.}(2005)Manchester, Hobbs, Teoh, \&
  Hobbs}]{manchesterAustraliaTelescopeNational2005a}
Manchester, R.~N., Hobbs, G.~B., Teoh, A., \& Hobbs, M. 2005, The Astronomical
  Journal, 129, 1993, \dodoi{10.1086/428488}

\bibitem[{Mardia(1975)}]{mardiaStatisticsDirectionalData1975}
Mardia, K.~V. 1975, Journal of the Royal Statistical Society: Series B
  (Methodological), 37, 349, \dodoi{10.1111/j.2517-6161.1975.tb01550.x}

\bibitem[{Prandoni {et~al.}(2017)Prandoni, Murgia, Tarchi, Burgay, Castangia,
  Egron, Govoni, Pellizzoni, Ricci, Righini, Bartolini, Casu, Corongiu,
  Iacolina, Melis, Nasir, Orlati, Perrodin, Poppi, Trois, Vacca, Zanichelli,
  Bachetti, Buttu, Comoretto, Concu, Fara, Gaudiomonte, Loi, Migoni, Orfei,
  Pilia, Bolli, Carretti, D'Amico, Guidetti, Loru, Massi, Pisanu, Porceddu,
  Ridolfi, Serra, Stanghellini, Tiburzi, Tingay, \&
  Valente}]{prandoniSardiniaRadioTelescope2017}
Prandoni, I., Murgia, M., Tarchi, A., {et~al.} 2017, A\&A, 608, A40,
  \dodoi{10.1051/0004-6361/201630243}

\bibitem[{Ransom(2011)}]{ransomPRESTOPulsaRExploration2011}
Ransom, S. 2011, Astrophysics Source Code Library, ascl:1107.017

\bibitem[{Ransom {et~al.}(2002)Ransom, Eikenberry, \&
  Middleditch}]{ransomFourierTechniquesVery2002}
Ransom, S.~M., Eikenberry, S.~S., \& Middleditch, J. 2002, astro-ph

\bibitem[{{van der Klis}(1989)}]{vanderklisFourierTechniquesXray1989}
{van der Klis}, M. 1989, in Timing {{Neutron Stars}}: Proceedings of the {{NATO
  Advanced Study Institute}} on {{Timing Neutron Stars}} Held {{April}} 4-15,
  27

\end{thebibliography}
\bibliographystyle{aasjournal}
\appendix

\section{Statistical properties of the binned Z statistic}
We verify here that the \zsqbin values agree closely with the expected \chizsq distribution in the case of white noise.

\begin{figure*}
\centering
\includegraphics[width=\textwidth]{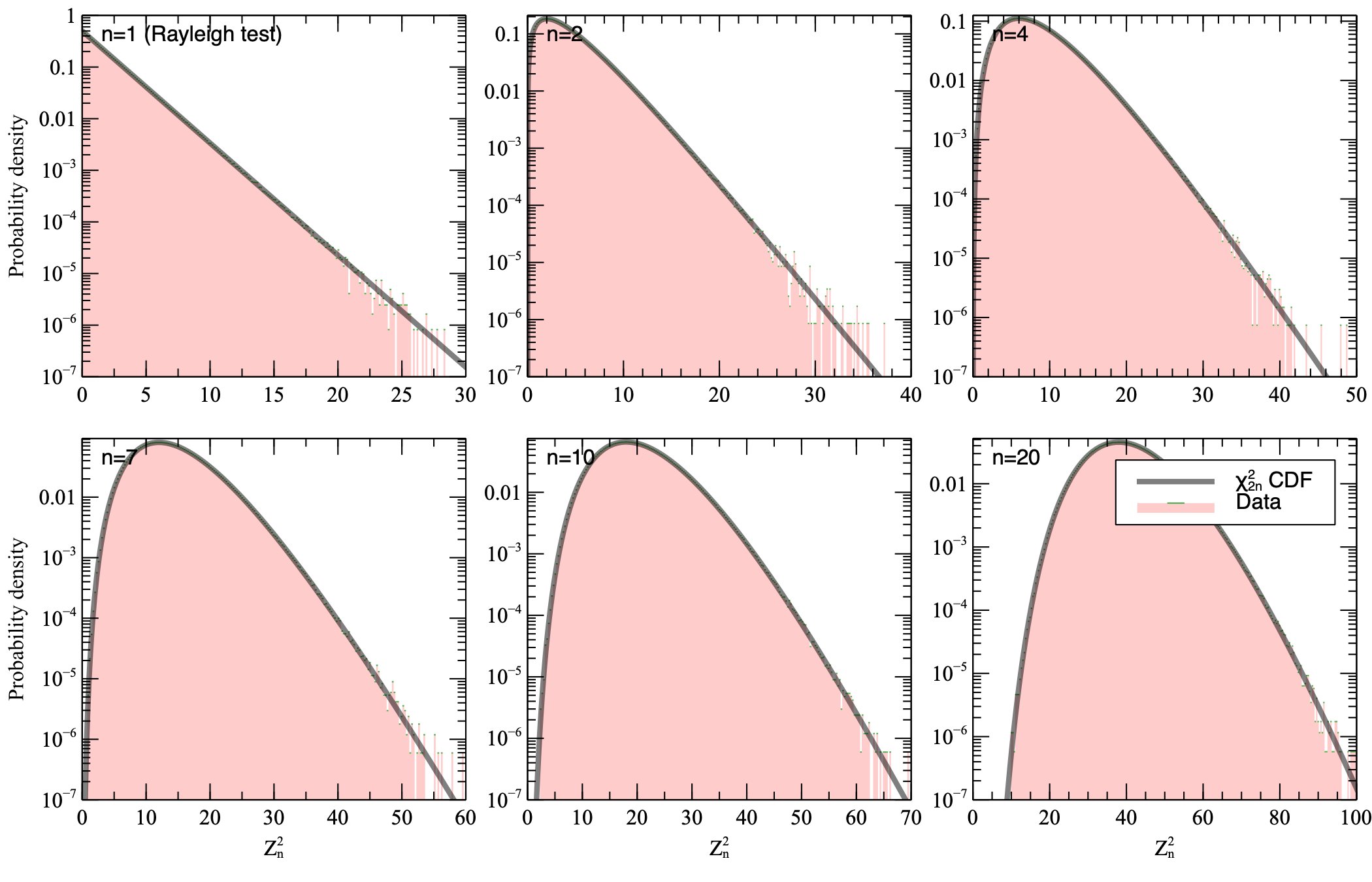}
\caption{Statistical distribution of the \zsq values for white noise data, with different values of $n$, compared to the expected \chizsq distributions.
We only selected values of \zsq obtained starting from pulsed profiles with least $2n$ bins (one needs at least two bins to contain a sinusoidal harmonic; this is analogous to the Nyquist theorem for the Fourier Transform).
\label{fig:znprob}}
\end{figure*}

To measure this agreement, we used 10,000,000 simulations of binned profiles with different numbers of bins (between 4 and 1024) and total number of photons (between 10 and 10,000).
For each value of $n$, we selected all simulations with $\nbin>2n$ (as per the Nyquist argument in \secref{sec:binned}).
Then, we compared the distribution of the \zsqbin values with the expected \chizsq distribution, in particular in the tails where the values are more important for significance estimates.
From this first test, it is clear that the approximation yields the expected probability distribution for \zsq for all values of $n$ if $\nbin> 2n$ (see \figref{fig:znprob}).

\section{What is the difference between the Z statistic for binned and unbinned event data?}\label{sec:difference}
The difference between the \zsq formulations in Equations~\ref{eq:zn} and~\ref{eq:znbin} is the fact that the phase of each event is approximated with the phase at the middle of a profile bin.
Looking at \eqref{eq:zn}, we see that it is of the form 

\begin{equation}\label{eq:zntransapp}
\zsq =\dfrac{2}{N} \sum_{k=1}^n \left[A_k^2 + B_k^2\right]
\end{equation}

where
\begin{eqnarray}
A_k = \sum_{j=1}^{N} \cos k\phi_j\\
B_k = \sum_{j=1}^{N} \sin k\phi_j
\end{eqnarray}
Now, let us express each phase as the sum of the phase at the center of each bin and an additional (usually small) term
\begin{equation}
    \phi_j = \phi_{b(j)} + \epsilon_j
\end{equation}

Using the standard trigonometric formulas for the cosine and sine of sums of angles, $A_k$ and $B_k$ become
\begin{eqnarray}
A_k = \sum_{b=1}^{\nbin} \left[\cos k\phi_b \sum_{i=1}^{w_b} \cos k \epsilon_i - \sin k\phi_b \sum_{i=1}^{w_b} \sin k \epsilon_i\right]\label{eq:A}\\
B_k = \sum_{b=1}^{\nbin} \left[\sin k\phi_b \sum_{i=1}^{w_b} \cos k \epsilon_i + \cos k\phi_b \sum_{i=1}^{w_b} \sin k \epsilon_i\right]\label{eq:B}
\end{eqnarray}

Let us now take a closer look at the inner sums. 
We are referring the $\epsilon_i$ to the center of the profile bins, so it is reasonable to assume that, in general, they will be equally distributed in a range $-\pi/\nbin \leq\epsilon_i \leq pi/\nbin$ around the bin centers\footnote{
This assumption breaks down for pulse profiles with very sharp features well above the noise level, but this implies high significance in any case}.
Therefore, 
\begin{equation}
    \sum_{i=1}^{w_b} \sin k \epsilon_i \approx 0
\end{equation}

and
\begin{eqnarray}
    \sum_{i=1}^{w_b} \cos k \epsilon_i &\approx& w_b \mathrm{mean}(\cos k \epsilon_i)\\
    &=& w_b \dfrac{1}{2\alpha_k}\int_{-\alpha_k}^{\alpha_k}\cos k \epsilon_i \,\mathrm{d}\epsilon_i
\end{eqnarray}
where 
$$
\alpha_k = \frac{\pi}{\nbin}
$$
so that
\begin{equation}
    \sum_{i=1}^{w_b} \cos k \epsilon_i \approx w_b \frac{\sin k\alpha_k}{k\alpha_k}.
\end{equation}

Concluding, Equations~\ref{eq:A} and~\ref{eq:B} become
\begin{eqnarray}
A \approx \frac{\sin k\alpha_k}{k\alpha_k} \sum_{b=1}^{\nbin} w_b \cos k\phi_b\\
B \approx \frac{\sin k\alpha_k}{k\alpha_k} \sum_{b=1}^{\nbin} w_b \sin k\phi_b
\end{eqnarray}
and \eqref{eq:zntransapp} becomes
\begin{equation}
    \zsq \approx 
    \frac{2}{N}\sum_{k=1}^n {\left(\frac{\sin k\alpha_k}{k\alpha_k} \right)}^2
    \left[
        {\left(\sum_{b=1}^{\nbin} w_b \cos k\phi_b\right)}^2 + 
        {\left(\sum_{b=1}^{\nbin} w_b \sin k\phi_b\right)}^2 
    \right]
\end{equation}

Comparing it to \eqref{eq:znbin}, we see that this formula only differs by the factor
\begin{eqnarray}
    C(k) &=& {\left(\frac{\sin k\alpha_k}{k\alpha_k} \right)}^2 \\
         &=& {\left(\frac{\nbin}{\pi k} \sin \frac{\pi k}{\nbin} \right)}^2\label{eq:correctionapp}
\end{eqnarray}
that goes to 1 for $k/\nbin \ll 1$, as expected, and drops to $(2/\pi)^2=0.405$ for $k=\nbin/2$ .

This is the \zsq equivalent of the loss of sensitivity of Fourier Power Density Spectra (PDS) at high frequencies due to sampling: a light curve of length $T$ sampled with $N$ bins is equivalent to convolving a continuous light curve with a ``binning window'', non-zero only between times $t-0.5\Delta t$ and $t+0.5\Delta t$, where $\Delta t=T/N$ is the sampling time. 
The Fourier transform of this sampling window is \citep[See eq. 2.19 from][]{vanderklisFourierTechniquesXray1989}
\begin{equation}
    B(f) = \frac{\sin \pi f T/ N}{ \pi f T/ N}
\end{equation}

and its square, similarly to \eqref{eq:correctionapp}, goes from a maximum of 1 at low frequencies to $(2/\pi)^2$ at the Nyquist frequency, acting as a low-pass filter.

\section{The binned \ef of a composition of sinusoidal signals}\label{sec:sumef}
Let us consider a pulsed profile composed of a sum of $l$ sinusoidal harmonics with random phases $\phi_{0, l}$.
Neglecting binning effects (i.e. choosing a number of bins $\nbin \gg m$), the profile can be described as:
\begin{equation}\label{eq:sinprof}
    p_i = \lambda \left[1 + \sum_{l=1}^m a_l \sin l(\phi_i - \phi_{0, l})\right]
\end{equation}

The \ef statistic in \eqref{eq:efbin} gives
\begin{eqnarray}
    \ef &=& \frac{1}{\lambda} \sum_{i=1}^{\nbin} {\left(p_i - \lambda\right)}^2\\
        &=& \frac{1}{\lambda} \sum_{i=1}^{\nbin} {\left[\sum_{l=1}^m a_l \sin l(\phi_i - \phi_{0,l})\right]}^2\\
        &=& \frac{1}{\lambda} \sum_{i=1}^{\nbin} {\left[\sum_{l=1}^m a_l \sin l\phi_i\cos l\phi_{0,l} - \sin l\phi_{0,l}\cos l\phi_i\right]}^2
\end{eqnarray}
We now manipulate the trigonometric functions and note that: 
the sum of any odd power of sinusoidal functions integrates to zero over a full cycle, so that only the terms with squares survive;
the sinusoidal harmonics form an orthogonal base, so that any integral of products of different harmonics integrates to zero as well. 
Then, we get

\begin{eqnarray}
    \ef &=& \frac{1}{\lambda} \sum_{l=1}^m a_l \left(\cos^2 l\phi_{0,l}\sum_{i=1}^{\nbin} \sin^2 l\phi_i + \sin^2 l\phi_{0,l} \sum_{i=1}^{\nbin} \cos^2 l\phi_i\right)
\end{eqnarray}

The integral of the square of a sinusoidal harmonic over a pulse profile gives $\nbin/2$, so it is easy to obtain
\begin{equation}\label{eq:ssig_appendix}
\ssig \approx \frac{\lambda\nbin}{2}\sum_{l=1}^{n} a_l^2 
\end{equation}

\section{The binned Z statistic of a composition of sinusoidal signals}\label{sec:sumzsq}

Again, we start from the profile in \eqref{eq:sinprof} and calculate \zsqbin.
We assume that $n\geq m$, i.e. the \zsqbin formula contains enough harmonics to describe the signal completely.
Also, $\nbin \gg n$, so that we can neglect the effect of binning.

As we did previously, we divide the \zsqbin formula in two terms
\begin{equation}\label{eq:zntrans}
\zsqbin =\dfrac{2}{\sum^i p_i} \sum_{k=1}^n \left[A_k^2 + B_k^2\right]
\end{equation}

where we use the profile bins $p_i$ as weights ($p_i$ in \eqref{eq:znbin}) and
\begin{eqnarray}
A_k &=& \lambda \sum_{i=1}^{\nbin} \cos k\phi_i \left[1 + \sum_l a_l\sin l(\phi_i - \phi_{0,l})\right]\\
B_k &=& \lambda \sum_{i=1}^{\nbin} \sin k\phi_i \left[1 + \sum_l a_l\sin l(\phi_i - \phi_{0,l})\right]
\end{eqnarray}
Applying the same arguments of Appendix~\ref{sec:sumef} (all sums of odd powers of sinusoidal functions over a full profile go to zero, orthogonality of sinusoidal harmonics), it is easy to get
\begin{eqnarray}
    A_k &=& \lambda\sum_i \cos k \phi_i \sum_l a_l \left(
                \sin l \phi_i\cos l\phi_{0,l} - 
                \cos l\phi_i\sin l \phi_{0,l} 
            \right) \nonumber\\
        &=& \lambda a_k \sum_i \cos k \phi_i \left(
                \sin k \phi_i\cos k\phi_{0,k} - 
                \cos k\phi_i \sin k \phi_{0,k} 
            \right) \nonumber\\
        &=& \lambda a_k 
            \left[
                \cos k\phi_{0,k} \sum_i \cos k \phi_i \sin k \phi_i - 
                \sin k\phi_{0,k} \sum_i \cos^2 k\phi_i 
            \right]\nonumber\\
        &=& -\frac{\nbin\lambda}{2} a_k \sin k \phi_{0,k}\\
    B_k &=& -\frac{\nbin\lambda}{2} a_k \cos k \phi_{0,k}
\end{eqnarray}

And finally
\begin{eqnarray}
\zsqbin &=& \frac{2}{\nbin\lambda} \sum_k{\left(
                \frac{\nbin\lambda a^2_k}{2}\right)}^2
                \left[
                    {\left(\sin k \phi_{0,k}\right)}^2 + 
                    {\left(\cos k \phi_{0,k}\right)}^2
                \right]\\
        &=&\frac{\nbin}{2\lambda}\sum_{k=1}^{n} a_k^2
\end{eqnarray}

\section{Inter-bin Pulse Profile Correlations due to Folding Methodology}\label{app:drizzle}
As discussed in \S\ref{sec:intro} near Eq.~\ref{eq:folding}, there is an alternative method to fold a time series that is arguably better when the time series bins are comprised of integrated samples or events.  In that case, instead of each time series bin being treated as a delta function in time, which can be placed at a specific phase in a pulse profile, each time series bin has a beginning and end in both time and in pulse phase.  The time series values are then spread proportionally over the corresponding pulse profile bins, effectively ``drizzling'' the integrated data over the corresponding portions of the accumulating pulse profile.  This is, in fact, how the program {\tt prepfold} from {\sc PRESTO} folds data.

Because the time series values are spread proportionally, neighboring bins of the pulse profile end up slightly correlated.  The amount of correlation depends on the ratio of the duration in phase of a pulse profile bin, $\Delta\phi_p = 1/\nbin$, to the duration of a time series bin $\Delta t$ in units of pulse phase, $\Delta\phi_t = \Delta t/P$, where $P$ is the folding period.  The ratio, which we will call
\begin{equation}
    \psi = \frac{\Delta\phi_p}{\Delta\phi_t} = \frac{P}{\nbin \Delta t},
\end{equation}

describes how much time resolution you have in your time series compared to the rapidity of the searched-for pulsations.

In the limit of small $\psi$ (i.e.~values near zero), the pulse profile bins become perfectly correlated (and will therefore never show any pulsations), whereas for large $\psi$, the pulse profile bins are basically uncorrelated and we effectively reproduce the folding methodology of Eq.~\ref{eq:folding}.  Much pulsar searching happens in the middle range where $\psi$ is between 1$-$100.

The result of these correlations is that the effective number of degrees of freedom \dof in the folded profile, for a $\chi^2$ sensitivity calculation, for instance, decreases below $\nbin-1$.  Alternatively, you can think of the effect as decreasing the variance in, or a smoothing of, the resulting pulse profile.

We have developed a semi-analytic correction $C$ (with help from Paul Demorest and Walter Brisken) to the relevant statistics due to the correlations of the form:
\begin{equation}
    C = a \psi \left(1 + \psi^b \right)^{(-1/b)},
    \label{eq:dofcorr}
\end{equation}

so that the effective number of degrees of freedom $\dof_\textrm{eff}$ to use in statistical tests is $\dof_\textrm{eff} = C \dof = C (\nbin - 1)$.

We performed a large number of simulations over a wide range of $\psi$, using time series of pure Gaussian noise and the folding code in {\sc PRESTO}, to determine both the validity of Eq.~\ref{eq:dofcorr} as well as the values of $a$ and $b$: $a = 0.96$ and $b = 1.806$.  The correction is good to a fractional error of less than a few percent as long as $\psi \gtrsim 0.5$.  There is a small dependence on \nbin which becomes apparent when $\psi \lesssim 0.7$.

To correct the measured noise level $\sigma_\textrm{meas}$ in the profile (for estimating a signal-to-noise ratio or flux density via the radiometer equation, for example), dividing by the square-root of $C$ will inflate $\sigma_\textrm{meas}$ appropriately: $\sigma_\textrm{corr} = \sigma_\textrm{meas} / \sqrt{C}$, where $\sigma_\textrm{corr}$ is the corrected standard deviation of the profile noise.

\end{document}